\documentstyle[12pt,psfig,amssymb,referee]{aa}

\def\EQ{\begin{equation}}
\def\EN{\end{equation}}
\def\EQA{\begin{eqnarray}}\def\ENA{\end{eqnarray}}

\begin{document}

\title{A powerful local shear instability in stratified disks}

\author{D. Richard \inst{1,2} \and F. Hersant \inst{1,3} \and O. Dauchot \inst{1} \and  F. Daviaud \inst{1} \and B. Dubrulle \inst{1,4} \and J-P. Zahn \inst{2}}

\institute{ GIT/SPEC/DRECAM, CEA Saclay, F-91190 Gif sur Yvette Cedex, France
\and  DASGAL, Observatoire de Paris, F-92190 Meudon, France
\and DESPA, Observatoire de Paris, F-92190 Meudon, France
\and CNRS, SNC 2005}

\authorrunning{Richard et al.}

 \date{Received ; accepted }

\abstract{
In this paper, we show that astrophysical accretion disks are
dynamically unstable to non-axisymmetric disturbances. This
instability is present in any  stably stratified
anticyclonically sheared flow as soon as the
angular velocity decreases outwards. In the large Froude number
limit, the maximal growth rate is proportional to the angular
rotation velocity, and is independent of the stratification. In the
low Froude number limit, it decreases like the inverse of the Froude
number, thereby vanishing for unstratified, centrigugally stable
flows. The instability is not sensitive to disk boundaries. We 
discuss the possible significance of our result,
and its implications on the turbulent state achieved by the disks.
We conclude that this linear instability is
one of the best candidates for the source of turbulence in
geometrically thin disks, and that magnetic fields can be safely 
ignored when studying their
turbulent state. The relevance of the instability
for thick disks or nearly neutrally stratified disks remains to be
explored.
     }

\maketitle

\section{Introduction}

The simplest model of an accretion disk is that of an axisymmetric
rotating shear flow in hydrostatic vertical equilibrium, with a
Keplerian velocity law. The
hydrostatic state generates a vertical stratification. If this
stratification is unstable, it leads to turbulence via convective
instability. When the stratification is stable, it is generally
ignored or thought to be unimportant in the stability analysis, under
the rationale that it can only {\sl stabilize} the flow. Ignoring the
stratification makes the accretion disk look like a simple
differentially rotating shear flow, with an azimuthal keplerian velocity
profile $V( r )\sim r^{-1/2}$ . Its linear stability with respect to
axisymmetric disturbances is governed by the Rayleigh criterion in the inviscid
limit:
\EQ
\frac{d(r V)^2}{dr}>0,
\label{rayleigh}
\EN
for stability.  Flows obeying this criterion are called {\sl
centrifugally stable}. The keplerian flow, in which angular momentum
increases outwards, falls into this category. Yet, there are 
observational evidences that astrophysical (putatively keplerian) 
disks are turbulent, and thus that a source of instability exists in 
these flows.\

Various mechanism have been found able to destabilize
centrifugally stable flows. They may or may not apply to
astrophysical disks.
\par i) Centrifugally stable flows can be destabilized  by 
non-axisymmetric instabilities  (Papaloizou \& Pringle 1984), but
these instabilities generally involve reflecting boundary conditions over sharp
edges, which appear unrealistic in the context of astrophysical
disks.\

\par ii) Centrifugally stable flows can also be destabilized by finite
amplitude disturbances involving a non-linear mechanism not captured
by the Rayleigh criterion (Dubrulle 1993; Richard \& Zahn 1999). 
Laboratory experiments
have shown that there is a critical Reynolds number above which the
flow becomes turbulent (Wendt 1933; Taylor 1936):
\EQ
R=\frac{r\Omega\Delta r}{\nu}>R_c.
\label{seuil}
\EN
Here, $\Delta r$ is the radial extent of the flow (the gap),
$\Omega=V/r$ is the rotation, and $\nu$ is the viscosity.
The critical Reynolds number increases with gap width. For  $\Delta r/r \longrightarrow 0$, it tends to that of plane Couette flow:
$R_c \approx 2,000$, and  
for large enough gap ($\Delta r/r > 1/20$), the instability criterion 
becomes :
\EQ
{r^3 \over \nu} {\Delta \Omega \over \Delta r} > R^*_c ;
\EN
this critical gradient Reynolds number $R^*_c$ is of order
$6\times 10^5$ when the inner cylinder is at rest, which is
the only case which has been thoroughly explored.

\par iii) Centrifugally stable flows can be further destabilized via a
physical {\sl catalyzer}, an additional ingredient making the flow
unstable at smaller Reynolds numbers. A first instance is a vertical
magnetic field (Chandrasekhar 1961).
This magnetic field provides a linear axi-symmetric instability
mechanism for centrifugally stable flows via an interchange
instability if $d(V/r)^2<0$, i.e. for anticyclonic sheared flows. The
application of this mechanism to disks was first discussed by
Balbus and Hawley (1991); they showed that  the stratification of the disk does
not modify the result, and  that the maximal growth rate of
instability in that case is proportional to the angular
rotation velocity in the inviscid limit. The influence of viscosity
and magnetic diffusivity $\eta$ has been recently
numerically studied by R\"udiger and Zhang (2001); they found that
instability occurs above
a critical Reynolds number, which depends on the magnetic Prandtl number
$P_m=\nu/\eta$. From their numerical simulations, they fit $R_c=51
Pm^{-0.65}$ down to $Pm=0.01$. In cold disks, $Pm=10^{-5}$, and so
$R_c$ is of the order $10^5$ (in the inner part of the disk)
(Rudiger \& Zhang 2001). For anyone familiar with MHD flows, the
instability of a centrifugally stable flow subject to vertical
magnetic field is a surprise: magnetic fields generally inhibit
instabilities, and thus, have a {\sl stabilizing} influence. In the
present case, it appears that a stabilizing factor acts upon a stable
flow so as to generate instability!

iv)
Very recently, Molemaker et al.
(2001) discovered that a similar catalyzing effect could 
be provided by a vertical stable
stratification, which induces a linear non-axisymmetric instability for all
anticyclonically sheared flow. Astrophysical disks are subject to
both vertical and radial stratification. Our primary goal here is to
show that this mechanism may apply to any accretion disk and that the
inclusion of the radial stratification essentially does not modify
the instability. Our second goal is to discuss the importance of this mechanism
in turbulence generation in disks, and compare it to other proposed
mechanisms, namely ii) and iii).

\section{A shearing instability in a stratified disk}

\subsection{Basic equations}
We consider a rotating compressible stratified disk of finite
vertical extent, with velocity ${\bf U}=r\Omega(r) e_\theta$ and
density and pressure  $\rho_0,P$. We consider perturbation of the
basic state ${\bf u},\rho, p$. In order to eliminate the acoustic waves from
the problem, we shall work in the Boussinesq approximation. The basic
dynamical equations ruling the perturbation are in this approximation:
\EQA
\nabla\cdot(\rho {\bf u})&=&0,\nonumber\\
D_t {\bf u} +{\bf u}\cdot \nabla {\bf U}+\frac{1}{\rho_0}\nabla p+\frac{3h}{5
\rho_0}\nabla P&=&0,\nonumber\\
D_t h+{\bf u}\cdot\nabla H&=&0.
\label{basicequation}
\ENA
Here, $D_t=\partial_t +{\bf U} \cdot \nabla$ is the Lagrangian derivative,
$H=\ln(P\rho_0^{-5/3})$  and $h=-5 \rho/3\rho_0$ are the entropy and
the entropy perturbation in the Boussinesq approximation, and assuming for 
simplicity the perfect gas law. Note that
the incompressibility condition can be used to eliminate the pressure
via the Poisson equation:

\EQ
\frac{1}{\rho_0}\Delta p=-\nabla\cdot\left(D_t u+u\nabla
U+\frac{3h}{5\rho_0}\nabla P\right).
\label{poisson}
\EN
We work in the local approximation, and decompose the perturbation
over Fourier modes with wavenumber $(k_r,k_\theta,k_z)$. Because of
the mean velocity, the perturbation azimuthal coordinate increases
with time like:
\EQ
\theta(t)=\theta(0)+k_\theta r\Omega t.
\label{travellingwave}
\EN
Because of the shear, the radial wavenumber increases with time
according to (Dubrulle \& Knobloch 1993)
\EQ
k_r(t)=k_r(0)-k_\theta S t
\label{shearedwave}
\EN
where $S=r\partial_{r} \Omega$ is the shear rate.
To take these two effects into account, we then decompose our
perturbations along a ``travelling" basis
$\exp[\omega t+i r k_\theta(\Omega-S) t+i {\bf k}\cdot {\bf r}]$, with
${\bf k}\cdot {\bf r}<<1$. To leading
order in $1/(k r)$ (i.e. considering that $S$ is independent of r),
and using cylindrical coordinates the Poisson equation becomes:
\EQ
-i k^2 \frac{\partial_r P}{\partial_z P}p=2 \Omega (k_\theta u-k_r
v)+s \left(k_r N_r N_z+k_z N_z^2\right),
\label{poissonlocal}
\EN
where $k^2=k_r^2+k_\theta^2+k_z^2$, ${\bf
u}=\partial_{z}P/\partial_{r} P(u,v,w)$ and $s=-h/\partial_{z} H$.
Here, we have introduced
the two components of the Brunt-Vaissala frequency:
\EQA
N_r^2&=&-\frac{3}{5\rho_0}\partial_r P\partial_r H,\nonumber\\
N_z^2&=&-\frac{3}{5\rho_0}\partial_z P\partial_z H,
\label{Bvdefinition}
\ENA
and used the assumption of rotation on cylinders $\Omega( r)$ :
\EQ
\partial_r P\partial_z H=\partial_z P\partial_r H.
\label{isobaric}
\EN
Note that in deriving (\ref{poissonlocal}) we have taken care to
first taking the divergence, and then replace the derivative in front
of $(u,v,w,p)$ by $ik(u,v,w,p)$.
Using (\ref{poissonlocal}), we obtain  the dynamic equation under the
local approximation as:
\EQA
k^2\omega u&=&2\Omega v\left(k^2-k_r^2\right)+2\Omega uk_r
k_\theta+s\left(N_r N_z(k^2-k_r^2)+k_r k_z N_z^2\right),\nonumber\\
k^2\omega v&=&-2\Omega vk_r k_\theta+2\Omega u( k_\theta^2-k^2)-S u
k^2+s\left(-N_r N_z k_r k_\theta+k_\theta k_z N_z^2\right),\nonumber\\
k^2\omega w&=&2\Omega v k_r k_z+2\Omega u k_z k_\theta+s\left(-N_r
N_z k_r k_z+(k_z^2- k^2) N_z^2\right),\nonumber\\
\omega s&=&-\frac{N_r}{N_z}u+w.
\label{basicequationlocal}
\ENA

To illustrate the basic mechanism of instability, it is convenient to
study first the case where $N_r^2=0$ (no radial stratification). We
then return to treat the more general case by building on the results
of this artificial example.

\subsection{Dispersion relation for $N_r^2=0$}
Setting the determinant of (\ref{basicequationlocal}) to zero, with
$N_r=0$ and $N_z\equiv N$, we find after some algebra and rearrangement:
\EQ
\omega^4 k^2 + \omega^2 \left(N^2(k_r^2+k_\theta^2)+ \kappa^2 k_z^2 
+2\Omega S k_\theta^2\right) +2\Omega S N^2 k_\theta^2 =0.
\label{dispers}
\EN
Here, $\kappa^2=4\Omega^2+2\Omega S$ is the epicyclic frequency.
In the axi-symmetric limit: $k_\theta=0$, and the
dispersion relation becomes:
\EQ
(k_r^2+k_z^2)\omega^4+ (N^2 k_r^2+k_z^2 \kappa^2)\omega^2 =0.
\label{dispersaxi}
\EN
There are two branches of solution: one with $\omega=0$ corresponding
to neutral modes propagating at the mean rotation speed. In the other
branch, $(k_r^2+k_z^2)\omega^2=-(k_r^2 N^2 + k_z^2 \kappa^2)$. In the
stably stratified, centrifugally stable case we consider, both $N^2$
and $\kappa^2$ are positive, and $\omega$ is imaginary. This branch
corresponds to gravito-inertial waves. As we shall now see, the linear
non-axisymmetric instability starts from the neutral branch of
solution.
In the general case (\ref{dispers}) is a quadratic equation in
$\omega^2$, which admits two roots,
    which may be real and distinct, real and equal, or complex conjugated. A
sufficient condition for instability is therefore that one of the
root has a positive real part. The root properties are determined by
the sign of $2\Omega S=\partial_r\Omega^2$. If this quantity is
negative, their product is negative, and the two roots are real of
opposite sign, implying instability. In the other case, the two roots
are equal or complex conjugated, and their real part is negative. A
sufficient condition for instability is therefore that
\EQ
\partial_r \Omega^2 <0 ,
\EN
which is satisfied by Keplerian rotation.
The growth rate $(\omega^2)^{1/2}$ can be found near the neutral branch, 
for vanishing
$\omega^2$:
\EQ
\omega^2=-\frac{2k_\theta^2 N^2\Omega S}{ k_\theta^2 (N^2+2\Omega
S)+k_z^2 \kappa^2}.
\label{nearbranch}
\EN
The behavior depends on the Froude number $F=\Omega/N$. For strong
stratification ($F\ll 1$), the growth rate is
$\omega_r=\sqrt{-2S\Omega}$, i.e. independent of the stratification.
The azimuthal  wavenumbers in that case scales like $k_r\sim 1/\Delta
r$, while the vertical wavenumber scales like $k_r/F$ (Yavneh et al. 2001).
For large Froude number, the growth rate is
\EQ
\omega_r=\sqrt{\frac{-2k_\theta^2 N^2\Omega S}{ 2k_\theta^2 \Omega
S+k_z^2 \kappa^2}}.
\label{autrelim}
\EN
Since in that case $k_\theta\sim k_z\sim k_r$ (Yavneh et al. 2001), 
the growth rate
decays like $1/F$, in agreement with the numerical computations
of Molemaker et al. (2001).\

In summary, we have found that for non-axi-symmetric
disturbances, the flow is linearly unstable, while for
axisymmetric disturbances, the flow remains stable.

\subsection{The general case: $N_r^2\neq 0$}
We consider now the general case. The general dispersion formula is then:
\EQ
k^2\omega^4+\omega^2 \left((k_z N_r+k_r N_z)^2 +k_z^2
\kappa^2+k_\theta^2(N_r^2+N_z^2+2\Omega S)\right)+2\Omega S N_z^2
k_\theta^2 =0.
\label{dispers2}
\EN
In the axisymmetric limit, one obtains a slightly different form of
(\ref{dispersaxi}):
\EQ
(k_r^2+k_z^2)\omega^4+\omega^2 \left((N_r k_z +k_r N_z)^2+k_z^2
\kappa^2\right)=0.
\label{dispersaxi2}
\EN
Here we find again a neutral branch of solutions, plus a solution of
stable waves in the centrifugally stable case.\

In the non-axisymmetric limit, the dispersion relation has exactly
the same structure as in the case with no radial stratification (with
the product of the root depending only on $S\Omega$), and therefore
the necessary condition for instability is unchanged. Moreover, in a
thin disk disk $N_r\sim (H/r) N_z<<N_z$, so that the expression of
the growth rates is amost unaffected by the radial stratification.

\subsection{Singular limits}
In the present letter, we have used a crude local approximation to
derive the stability criterion. In fact, these kinds of local
approximations have known pitfalls (Dubrulle \& Knobloch 1993). In the present
case, it suggests for example that the growth rate of the
perturbation vanishes algebraically fast in $S$ and is independent
on the stratification. A more careful analysis of the normal mode
problem however shows that in the small gap limit, and for strong
stratification ($F<<1)$, the growth rate of the most unstable mode is
(Molemaker et al. 2001)
\EQ
\omega\approx -2 S  ( k_\theta r ) e^{4 \Omega /S}.
\label{careful}
\EN
Our analysis and the numerical
results of Molemaker et al. (2001) also shows that for large values 
of $F$, the
growth rate behaves linearly with $1/F$, thereby vanishing with
vanishing stratification. This shows that there is no singular limit
in this problem.\

\subsection{Non-linear saturation and implications for turbulence}

The present analysis was performed in the inviscid limit. The
influence of viscosity and the non-linear saturation of the
instability have been studied numerically by Molemaker et al. (2001) 
and by Yavneh et al. (2001). They
found that a small viscosity essentially does not change the results,
and only introduces a critical Reynolds number, above which the flow is
unstable. For example, at $F=0.01$, $Ro=-2/3$, the critical Reynolds
number is of order 1200, and from Section 2.2 
we may extrapolate this result to higher Froude number: $R_c=
1200 \sqrt{1+F^2}$. In thin astrophysical disks, we have $N_z^2\sim
\Omega^2 (\nabla_{\rm ad}-\nabla)$, meaning that the Froude number is
of order 10, if we assume a subadiabatic gradient of 1/10; therefore
$R_c \sim 10^4$.

The transition to turbulence then
proceeds via successive period doubling bifurcations (Molemaker et al. 2001).
Without stratification (and without magnetic field), the turbulence
would set in via finite amplitude disturbances and non-linear 
mechanisms, a very intermittent scenario, in which turbulent domains
progressively grow in size with increasing Reynolds number, until
they finally invade the whole flow, as observed in plane
Couette flow (Dauchot \& Daviaud 1994).
Therefore, an interesting open question is
what happens in the highly turbulent regime, and especially
whether this turbulence, triggered by a {\sl catalyzer}, resembles
the turbulence generated by non-linear instabilities. 

This question is important, because it is generally believed that
turbulent transport properties depend on the source of the
turbulence. The main argument is that to reach a stationary state,
turbulence must organize itself so as to suppress the cause of
instability. A good example is given by centrifugally unstable flows,
where the turbulent state is characterized by a flat angular momentum
distribution (outside boundary layers). But this holds only for
moderate Reynolds number flow (say up to $R=10^5$).
For larger Reynolds number, there is experimental evidence in the
rotating shear flow that a transition occurs above which the flow
presents all the characterisitics of classical shear (boundary layer)
turbulence (Lathrop et al. 1992). In astrophysical disks, the Reynolds
number tends to be much larger than this critical value, and one may
wonder whether the turbulent transport in disks is not universal.
It would therefore be very interesting to conduct
laboratory and numerically experiments of centrifugally stable,
stratified flows to
study that large Reynolds number turbulent regime.

\section{Conclusion}
We have shown that all accretion disks are subject to a
powerful, non-axisymmetric instability. This instability lurks in any
astrophysical disk, because of the stratification induced by
the vertical component of gravity. It is a purely hydrodynamical
instability and does not require the presence of any magnetic field,
whatever small. Therefore there is no reason to invoke
a ``dead zone'' in insufficiently ionized disks, as was done by Gammie (1996). 

In astrophysical disks, the critical Reynolds
number to trigger this instability is of the order of $10^4$, which is
less than the critical Reynolds number for both the magneto-rotational
instability and the finite-amplitude hydrodynamic
instability.  But the question of which instability occurs first has 
little relevance: what one should ask is what kind of turbulence develops
at high Reynolds number. In particular, it is not clear that
magnetic field plays an important role in that regime, because
in most turbulent flows studied so far, the level of magnetic energy
turned out to be an order of magnitude less than that of kinetic energy.
Of course, this does not rule out the transport of angular
momentum by a large scale magnetic field.

\end{document}